\documentclass[aps,prd,twocolumn,reprint]{revtex4-2}

\usepackage{graphicx}
\usepackage{bm}
\usepackage{hyperref}
\usepackage{amsfonts}
\usepackage{tikz}
\usepackage{amssymb}
\usepackage{graphicx}
\usepackage{color}
\usepackage{epsfig}
\usepackage{remreset}
\usepackage{amsmath,amsthm,ulem}
\usepackage{mathtools}
\usepackage{mathrsfs}
\usepackage{color}
\usepackage{float}
\usepackage{tensor}
\usepackage{lipsum}
\usepackage[caption=false]{subfig}
\usepackage{slashed}
\usepackage{lmodern}
\usepackage[T1]{fontenc}
\usepackage{pgf,pgfarrows,pgfnodes}

\begin{document}

\title{The bound orbits and gravitational waveforms of timelike particles around renormalization group improved Kerr black holes}

\author{Yong-Zhuang Li}
\email{liyongzhuang@just.edu.cn}
\affiliation{%
School of Science, Jiangsu University of Science and Technology, Zhenjiang 212100, Chian
}%

\author{Xiao-Mei Kuang}%
 \email{xmeikuang@yzu.edu.cn}
\affiliation{ 
Center for Gravitation and Cosmology, College of Physical Science and Technology, Yangzhou University, Yangzhou 225009, China
}%

\date{\today}

\begin{abstract}
 In this article, we investigate the bound orbits of the timelike particles and the gravitational waveforms emitted from these orbits around a renormalization group improved Kerr black hole in the framework of the asymptotic safety approach. The running Newton coupling in the metric is characterized by two free quantum parameters $(\omega,\,\gamma)$ arising from the non-perturbative renormalization group theory and the appropriate cutoff identification, respectively. As expected, the radii of the horizon, the marginally bound orbits and the innermost stable orbit are all decrease as the quantum parameters increase. Under the extreme mass-ratio inspirals approximation the deviation of gravitational waveforms radiated by the periodic orbits from those in the classical Kerr background increases with the two quantum parameters. However, this effect is much smaller in the retrograde case compared to the prograde case. Especially, by comparing the characteristic strain of those gravitational wave with the sensitivity curve of several potential detectors, we find that their characteristic frequencies can fall within the sensitivity ranges of several planned gravitational wave observatories, suggesting that such signals may be detectable with sufficient  instrumental sensitivity.
\end{abstract}

\maketitle

\section{Introduction}\label{sec:intro}

Born in the early twentieth century, General Relativity (GR) and Quantum Mechanics (QM) are arguably two of the most profound physical theories of all times. The subsequent evolution of quantum mechanics into the Standard Model, in particular, has successfully unified the electromagnetic, weak, and strong interactions. Nevertheless, efforts to quantize gravity have never ceased from the beginnings but have never truly succeeded. One crucial point is that GR is perturbatively non-renormalizable and thus is considered merely as an effective field theory, making the development of a fully consistent theory of quantum gravity difficult. Various possibilities have been investigated by physicists to formulate the ultraviolet complete quantum gravity, such as string theory \cite{Polchinski:1998rq,Polchinski:1998rr,Green:2012oqa,Green:2012pqa,Kiritsis:2019npv}, loop quantum gravity \cite{Rovelli:1997yv,Ashtekar:2017yom,Gambini:2020ypi,Ashtekar:2021kfp}, noncommutative
geometry\cite{Plauschinn:2018wbo,Marcolli:2018uea,Kar:2025baw}, Causal set theory \cite{Sorkin:2003bx,Surya:2019ndm}, see \cite{Bambi:2023jiz,Basile:2024oms,Buoninfante:2024yth} for more comprehensive introduction to different theories of quantum gravity.

Among these theories, the asymptotic safety approach to gravity (ASG hereafter) aims to achieve a consistent and predictive description of gravitational interactions within the framework of quantum field theory by finding a non-perturbative ultraviolet (UV) completion \cite{Reuter:1996cp,Niedermaier:2006wt,Ambjorn:2012jv,Hossenfelder:2012jw,Reuter:2019byg,Bonanno:2020bil,Basile:2025zjc}. Within such a paradigm, the fundamental gravitational degrees of freedom are postulated to be faithfully represented by the spacetime metric even in the trans-Planckian regime \cite{Weinberg:1980gg,Reuter:1996cp}. Under this assumption, the functional renormalization group flow of the effective average action $\Gamma_{k}$ exhibits a non-Gaussian UV fixed point-commonly referred to as the Reuter fixed point-in the infinite-dimensional theory space of diffeomorphism-invariant functionals \cite{Kawai:1993mb,Souma:1999at,Litim:2003vp}. The existence of this fixed point, characterized by a finite number of relevant (UV-attractive) directions, ensures the theory is UV-complete and predictive. As a result, quantum fluctuations are controlled in the high-energy limit, rendering the theory non-perturbatively renormalizable, and allowing for a consistent, unitary, and background-independent description of quantum spacetime across all energy scales \cite{Niedermaier:2006ns,Reichert:2020mja,Bambi:2023jiz}.
In practice, various approaches exist to constrain these theories, from entangled atom interferometers \cite{Marletto:2017kzi,Bose:2017nin,Krisnanda:2019glc,Carney:2021yfw}, high energy cosmological observations \cite{Cai:2014hja,Terzic:2021rlx,Cicoli:2023opf,Banerjee:2023uxk,Li:2025cxn}, to black hole shadow and gravitational waves \cite{Tarrant:2018mzs,Calcagni:2019ngc,Verlinde:2019xfb,Vermeulen:2020djm,Agullo:2020hxe,Guerreiro:2021qgk,Zurek:2022xzl,Vagnozzi:2022moj,Ayzenberg:2023hfw,Staelens:2023jgr,Li:2023djs,Yunes:2024lzm,Deppe:2025pvd} and so on. 

In this paper, we will primarily focus on the orbital motions of timelike particles outside a Kerr-like black hole in the ASG framework and the characteristics of gravitational waves (GW) emitted by periodic orbits, and evaluate the potential detectability of the corresponding GW by matching their characteristic strain with the sensitivity curve of several potential detectors, including LISA \cite{Robson:2018ifk}, eLISA \cite{Amaro-Seoane:2012vvq}, TianQin \cite{TianQin:2015yph,Torres-Orjuela:2023hfd}, LIGO \cite{Essick:2025zed,LIGOScientific:2025slb,ligo_scientific}, ASTROD-GW \cite{Ni:2012eh}, DECIGO \cite{Ishikawa:2020hlo,Iwaguchi:2020cxa}, TaiJi \cite{Hu:2017mde,Luo:2019zal,Gong:2021gvw,Liu:2023qap}, SKA $\&$ IPTA \cite{Hobbs2010CQGra}.

Our studies are motivated by two aspects. On one hand, black holes are natural laboratories to test gravity in the strong field regime. Particularly since the release of the black hole images (i.e., M87 and Sgr A*) and gravitational wave signals \cite{LIGOScientific:2016aoc,LIGOScientific:2018mvr,EventHorizonTelescope:2019dse,LIGOScientific:2020aai,EventHorizonTelescope:2022wkp}, theoretical and observational research into the properties of black hole vicinities has been thriving with unprecedented momentum. Especially, among the most promising sources for future space-based GW detectors are extreme mass ratio inspiral (EMRIs), in which a stellar mass compact object gradually spirals into a supermassive black hole \cite{Hughes:2000ssa,Glampedakis:2005hs}. The gravitational waveforms emitted from EMRIs encode rich information about the features of the central objects, such that this system is very important and provides a platform for testing GR or possible deviation because of the quantum effect and other new physics. A significant feature of EMRI dynamic is the construction from periodic orbits, which are bound trajectories of particles returning to its initial stater after completing an integer number of radial and angular oscillations \cite{Levin:2008mq}. The gravitational waveforms generated by EMRIs inherit distinctive imprints of such periodic orbits, characterized by zoom-whirl phases that occur during the inspiral \cite{Babak:2006uv,Poisson:2014kt,Maselli:2021men,Liang:2022gdk}, which has inspired extensive research on GW signals from periodic orbits in modified gravity and their potential detectability by future observations \cite{Tu:2023xab,Foschini:2024tgb,Li:2024tld,Haroon:2025rzx,Yang:2024lmj,Zhao:2024exh,Junior:2024tmi,Guo:2025scs,Choudhury:2025qsh,Chen:2025aqh,Haroon:2025rzx}.

On the other hand, the asymptotic-safety-inspired black-hole models has already constructed via the so-called gravitational renormalization group improvement (RGI). To put it simply, starting from a classical action, one can replace the gravitational couplings with the running counterparts, and then replace the running couplings with the solutions of the corresponding renormalization group (RG) equations complemented by suitable physical initial conditions. Finally, the RG scale parameterizing the running couplings is identified with a scale of the system which could act as a physical infrared (IR) cutoff. Thus, the properties of asymptotic-safety-inspired black hole spacetime can then be characterized by a set of corresponding additional parameters. We refer the readers to Refs. \cite{Platania:2019kyx,Eichhorn:2022bgu,Borissova:2022mgd,Platania2023}, which provide a comprehensive and detailed overview of the development of black holes within the ASG framework. In particular, with the given metrics of the RG-improved black holes, investigations have been done to identify quantum gravity modifications to the shadows, quasinormal modes and gravitational waves, as we elaborate below.

 This line of investigation was initially pursued by Held et al. in Ref. \cite{Held:2019xde}, where they examined modifications to both spinning and non-spinning black hole spacetimes that incorporate a scale-dependent Newtonian coupling. This scale dependence generated by quantum fluctuations is characterized by a single parameter, i.e. the square of the inverse transition scale to the fixed-point regime measured in Planck units. Unlike this, Lu and Xie considered a RGI-Schwarzschild black hole with a lapse function being corrected with two quantum parameters, and investigated the weak and strong deflection gravitational lensing \cite{Lu:2019ush}. This work has been generalized to the case with presence of plasma medium \cite{Atamurotov:2022iwj}. The dynamics, including the chaotic dynamics, of neutral, electrically charged and magnetized particles around RGI-Schwarzschild were studied in Ref. \cite{Rayimbaev:2020jye,Lin:2022llz,Lu:2024srb}. Shi and Cheng researched the shadow and gamma-ray bursts of a asymptotic-safety-inspired Schwarzschild black hole in the IR limit \cite{Shi:2023kid}. Using the Hamilton-Jacobi equation and Carter separable method, Kumar et al. analytically investigated the shadows cast by rotating black holes in the ASG by deriving complete null geodesics and observables, where the running Newtonian coupling depends on two ASG parameters \cite{Kumar:2019ohr}. Eichhorn and Held also constructed a novel family of regular black-hole spacetimes based on a locality principle which ties new physics to local curvature scales in Refs. \cite{Eichhorn:2021iwq,Eichhorn:2022bbn}, and explored the image features of the spinning black holes with disks. Their regular black holes can be connected to the ASG ones by setting the specified parameter. The shadow of a RGI rotating black hole has also been presented in Ref. \cite{Sanchez:2024sdm}, with a rotating metric obtained from the generalized Newman-Janis (NJ) algorithm, i.e., the running Newton coupling is related to a single quantum parameter. The circular orbits of a spinning test particle moving
in the equatorial plane of such a spinning spacetime were studied using the Mathisson-Papapetrou-Dixon equations together with the Tulczyjew spin-supplementary condition in Ref. \cite{Ladino:2022aja}. {Recently, Chen et al.  also suggested a new quantum improved Kerr black hole in ASG framework with an alternative definition of Newton coupling from the perspective of satisfying the zeroth and first laws of thermodynamics at the horizon  \cite{Chen:2022xjk,Chen:2023wdg}, of which the characteristics of black hole shadow images based on this definition were also discussed in Ref. \cite{Cao:2024vtq}.} For quasinormal modes and gray-body factors of asymptotic-safety-inspired black holes, see Refs. \cite{Konoplya:2022hll,Dubinsky:2024aeu,Lutfuoglu:2025ohb}. Especially, the links between the deformation parameter of the generalized uncertainty principle (GUP) to the two free parameters of the running Newtonian coupling constant of the ASG program has been conducted in Ref. \cite{Lambiase:2022xde}, and been tested by Lambiase et al. by calculating and examining the shadow and quasinormal modes of black holes \cite{Lambiase:2023hng}. However, few works have discussed the gravitational waves of the black hole systems under the ASG framework. The latest investigation is presented in Ref. \cite{Huang:2025czx}, where the gravitational waveforms generated by the different periodic orbits of timelike test particles around scale-dependent Planck stars or RGI-Schwarzschild black holes are investigated using both time-domain and frequency-domain methods.

The paper is organized as follows. In Sec. \ref{sec:background} we briefly review the properties of the RGI-Kerr black hole, then we investigate the timelike geodesics of the particles in Sec. \ref{sec:timelike}, mainly focusing on the precessing and periodic orbits. Based on the periodic orbits we examine the gravitational wave radiations in one
complete period of a test object {and evaluate their potential detectability} in Sec. \ref{sec:GW}, and finally give summaries and remarks in Sec. \ref{sec:conclusion}. 

\section{BACKGROUND}\label{sec:background}

To proceed, we shall briefly review the RGI-Kerr spacetime in the ASG framework. One approach is to simply replace the classical Newton's constant $G$ in the classical Kerr metric with the running one $G(r)$ \cite{Reuter:2010xb,Haroon:2017opl}, while another method involves deriving it from a static spherically symmetric solution via the Newman-Janis algorithm \cite{Bonanno:2000ep,Torres:2017gix}. {To construct $G(r)$ in the ASG theory, one needs to use the so-called effective average action (EAA), and truncate the renormalization group flow in the infinite dimensional space of all action functionals, known as “Einstein-Hilbert truncation”. Then using the background gauge formalism with a background metric $\bar{g}_{\mu\nu}$, the EAA has the form \cite{Reuter:1996cp,Bonanno:2000ep,Lauscher:2001ya,Torres:2017gix}
\begin{eqnarray}
    \Gamma_{k}[g,\bar{g}]&=&S_{EH}+S_{gf}[g,\bar{g}]\nonumber\\
    S_{EH}&=&\frac{1}{16\pi G(k)}\int d^{d}x\sqrt{g}\left[-\mathcal{R}(g)+2\bar{\lambda}(k)\right],
\end{eqnarray}
where $k$ is a mass-scale parameter. $G(k)$ and $\bar{\lambda}(k)$ denote the running Newton constant and cosmological constant, respectively. $S_{gf}[g,\bar{g}]$ is the classical background gauge fixing term. The $k$-dependence of the EAA is described by the flow equation
\begin{eqnarray}
    \partial_{t}\Gamma_{k}&=&\frac{1}{2}\text{Tr}\left[\left(k^{-2}\Gamma_{k}^{(2)}[g,\bar{g}]+\mathcal{R}_{k}^{gr}[\bar{g}]\right)^{-1}\partial_{t}\mathcal{R}_{k}^{gr}[\bar{g}]\right]\nonumber \\
    &\quad&-\text{Tr}\left[\left(-\mathcal{M}[g,\bar{g}]+\mathcal{R}_{k}^{gh}[\bar{g}]\right)^{-1}\partial_{t}\mathcal{R}_{k}^{gh}[\bar{g}]\right],
\end{eqnarray}
where $t\equiv\ln k$, and $\Gamma_{k}^{2}$ stands for the Hessian of $\Gamma_{k}$ with respective to $g_{\mu\nu}$. $\mathcal{M}$ is the Faddeev-Popov ghost operator. The operators $\mathcal{R}_{k}^{gr}$ and $\mathcal{R}_{k}^{gh}$ implement the IR cutoff in the graviton and the ghost sector. Inserting the EAA into the flow equation one will get a coupled system of differential equations for the Newton constant and the cosmological constant, while for former such equation is simplified as 
\begin{eqnarray}
    \partial_{t}g(t)=\beta(g(t)),
\end{eqnarray}
where $g(k)\equiv k^{d-2}G(k)$. The fixed points are then defined by $\beta(g(t))=0$. With the approximation $\bar{\lambda}\approx0$, solving the differential equation leads to 
\begin{eqnarray}
    G(k)=\frac{G_0}{1+\tilde{\omega} G_0 k^2},
\end{eqnarray}
where $\tilde{\omega}$ is an integration constant. Considering the black hole spacetime, $k$ can be converted to a position dependent quantity as 
\begin{eqnarray}
    k(P)=\frac{\xi}{d(P)},
\end{eqnarray}
where $\xi$ is a numerical constant and $d(P)$ is defined as the proper distance from the spacetime point $P$ to the center of the black hole along a curve $\mathcal{C}$ as
\begin{eqnarray}
    d(P)=\int_{\mathcal{C}}\sqrt{|ds^2|}.
\end{eqnarray}
For spherically symmetric spacetimes with Schwarzschild coordinates $(t,\, r,\,\theta,\,\phi)$, one will have 
\begin{eqnarray}
    d(r)=\left(\frac{r^3}{r+\gamma G_0M}\right)^{\frac{1}{2}},
\end{eqnarray}
where $\gamma$ can be treated as a free parameter and $M$ is the black hole mass. Finally, the running gravitational coupling constant is
\begin{equation}\label{eq:runG}
    G(r)=\frac{G_0 r^3}{r^3+\omega G_{0}\left(r+\gamma G_{0}M\right)},
\end{equation}
with $\omega=\tilde{\omega}\xi^2$.}

Using the definition (\ref{eq:runG}), the RGI-Kerr black hole metric has the following form in Boyer-Lindquist coordinates (with $c=1$)
\begin{eqnarray}\label{eq:metric}
    ds^{2}=&-&\left(1-\frac{2G(r)Mr}{\Sigma}\right)dt^{2}-\frac{4aG(r)Mr\sin^{2}{\theta}}{\Sigma}dtd\phi\nonumber\\
    &+&\frac{\Sigma}{\Delta}dr^{2}+\Sigma d\theta^{2} \nonumber \\
    &+&\sin^{2}{\theta}\left[r^2+a^2+\frac{2a^{2}G(r)Mr}{\Sigma}\sin^{2}{\theta}\right]d\phi^2,
\end{eqnarray}
where $G_0$ represents the well-known Newton's gravitational constant, and
\begin{eqnarray}
    \Sigma&=&r^2+a^2\cos^{2}{\theta},\\
    \Delta&=&r^2-2G(r)Mr+a^2.
\end{eqnarray}
Here, $\omega$ and $\gamma$ are parameters arsing from the non-perturbative renormalization group theory and the appropriate cutoff identification, respectively. Clearly, $\omega$ symbolizes the quantum corrections on the classical spacetime geometry, and the metric will return to the Kerr metric when $\omega=0$. The preferred theoretical values of $\gamma$ and $\omega$ have been discussed \cite{Donoghue:1993eb,Hamber:1995cq,Torres:2017gix}. However, we will consider them as positive free parameters. Meanwhile, the natural units are adopted throughout the entire article, i.e., $c=G_{0}=\hbar=1$ unless specified, and all numerical calculations are done in the units of $M=1$. 

The horizon of the RGI-Kerr black hole is determined by $\Delta=0$, which consequently establishes a constraint relation between $\omega$ and $\gamma$. In fact, the condition $\Delta=0$ indicates that
\begin{equation}\label{eq:omega}
    \omega=-\frac{r^{3}(r-r^{K}_{-})(r-r^{K}_{+})}{(r^{2}+a^2)(r+\gamma)},
\end{equation}
where $r^{K}_{\pm}=M\pm\sqrt{M^2-a^2}$ are the horizons of the Kerr metric. Since we have constrained $\omega>0$ and $\gamma>0$, so if we treat $\omega$ as a function of $r$ with $\gamma$ held fixed, then one has $\omega_{min}=\omega(0)=0$ and $\omega_{max}=\omega(r_0)$, where $r_0\in[r^{K}_{-},r^{K}_{+}] $ is one of the solutions to $\omega'(r)=0$. As long as $\omega\in[0,\omega_{max}]$, the horizon exists. Especially, the RGI-Kerr metric becomes extremal when $\omega=\omega_{max}$.

\begin{figure}
    \centering
    \includegraphics[width=\linewidth]{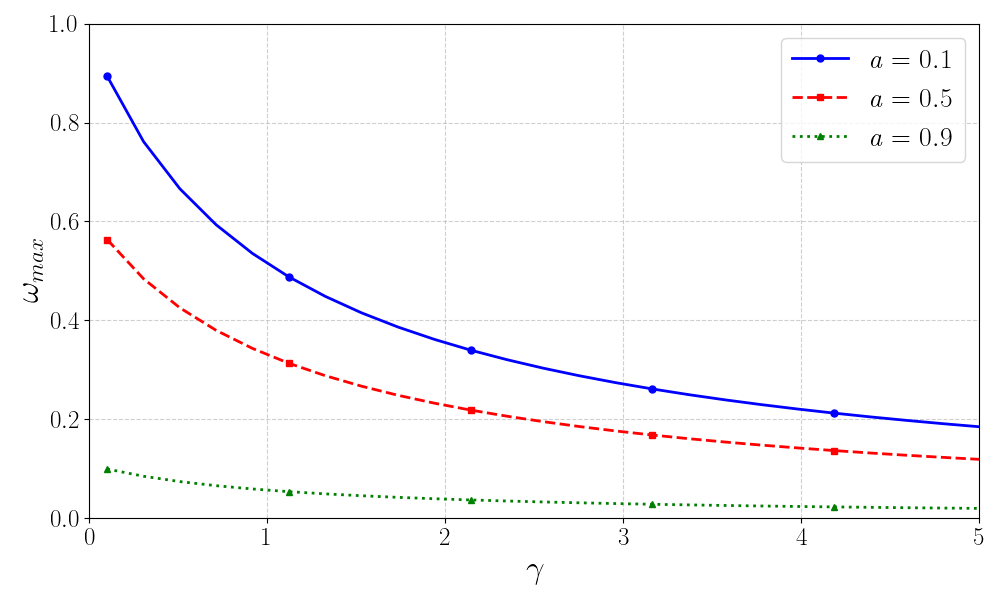}
    \caption{The maximum $\omega_{max}$ as a function of $\gamma$ with the spin parameter $a=0.1$ (blue solid), $a=0.5$ (red dashed) and $a=0.9$ (green dotted), respectively. The horizon exists in the parameter region $\omega\in[0,\omega_{max}]$ for fixed $\gamma$.}
    \label{fig:omegavsgamma}
\end{figure}

\begin{figure}
    \centering
    \includegraphics[width=\linewidth]{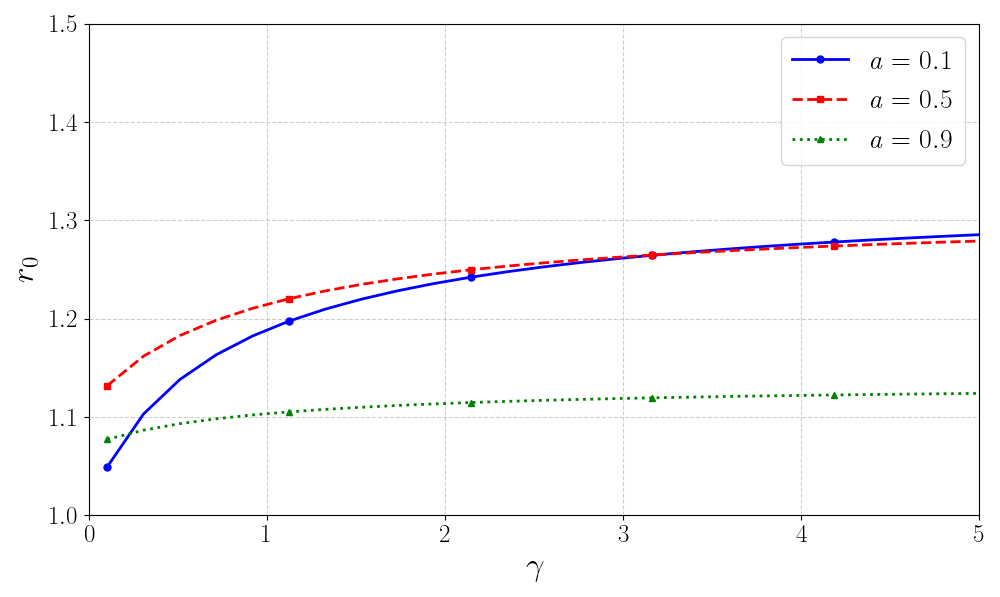}
    \caption{The horizon radius $r_0$ of the extremal RGI-Kerr black hole as a function of $\gamma$ with the spin parameter $a=0.1$ (blue solid), $a=0.5$ (red dashed) and $a=0.9$ (green dotted), respectively.}
    \label{fig:r0vsgamma}
\end{figure}

In Fig. \ref{fig:omegavsgamma} we numerically calculate the relation between $\gamma$ and $\omega_{max}$ for selected spin parameters $a=(0.1,\,0.5,\,0.9)$, which is consistent with the result shown in Ref. \cite{Xie:2024nqf}. In deed, it is evident that the numerator of left side of Eq. (\ref{eq:omega}) constitutes a concave function independent in $\gamma$, which means it has a fixed curve over the interval $[r_{-}^K,r_{+}^{K}]$. The denominator, on the other hand, represents a monotonically increasing function exhibiting linear dependence on $\gamma$. Thus, the monotonic growth of the denominator with increasing $\gamma$ unequivocally leads to a corresponding decrease in the ratio $\omega$, in complete agreement with the behavior demonstrated in Fig. \ref{fig:omegavsgamma}. A similar analysis can be directly extended to examine the relationship between $\omega_{max}$ and $a$. It is straightforward to verify that the extremum of the numerator decreases monotonically with increasing $a$, while $r_{-}^{K}$ expands and $r_{+}^{K}$ shrinks. Consequently, the numerator decreases while the denominator increases with growing $a$, which necessarily implies that $\omega_{max}$ must decrease as $a$ increases. This result demonstrates that as $a$ increases, the spacetime geometry of the RGI-Kerr black hole converges toward that of the classical Kerr solution.

Unfortunately, it is virtually impossible to analytically determine the effects of both $\gamma$ and $a$ on the horizon radius $r_0$ of the extremal RGI-Kerr black holes, leaving numerical approaches as the only viable alternative. The results demonstrate that for fixed $a$, $r_0$ increases monotonically with growing $\gamma$. In the large $\gamma$ regime, $r_0$ decreases monotonically as $a$ increases, while for sufficiently small $\gamma$ the extremal horizon radius $r_0$ initially increases then decreases with increasing $a$, see Fig. \ref{fig:r0vsgamma}. The radius of the event horizon for the non-extremal RGI-Kerr black hole depending on the parameters $\omega$ and $\gamma$ for different values of the black hole’s spin $a$ can be found in Ref. \cite{Ladino:2022aja}, suggesting that the increase of these parameters always produce a decrease in the radius of the outer horizon. In fact, the definition (\ref{eq:runG}) implies that the inequality $G(r) < G_0$ holds universally for all positive values of ${\omega}$ and $\gamma$. Consequently, the effective mass $G(r)M$ of the RGI-Kerr black hole is inherently smaller than the conventional Kerr black hole mass $G_0M$. Given that the outer horizon radius of a classical Kerr black hole is determined by $r^K_+ = M + \sqrt{M^2 - a^2}$, this mass reduction necessarily results in a contraction of the horizon radius.

\section{PRECESSING AND PERIODIC ORBITS OF THE TIMELIKE PARTICLES}\label{sec:timelike}

We now in this section derive the geodesic motion for a timelike particle in the RGI-Kerr spacetime. The Lagrangian is then written as
\begin{eqnarray}
    2\mathscr{L}=&-&\left(1-\frac{2G(r)Mr}{\Sigma}\right)\dot{t}^{2}-\frac{4aG(r)Mr\sin^{2}{\theta}}{\Sigma}\dot{t}\dot{\phi}\nonumber\\
    &+&\frac{\Sigma}{\Delta}\dot{r}^{2}+\Sigma \dot{\theta}^{2} \nonumber \\
    &+&\sin^{2}{\theta}\left[r^2+a^2+\frac{2a^{2}G(r)Mr}{\Sigma}\sin^{2}{\theta}\right]\dot{\phi}^2,
\end{eqnarray}    
where the dot denotes the derivative with respect to the affine parameter $\tau$. Considering that the quantum correction term appears solely in $G(r)$, the geodesic equations in RGI-Kerr spacetime should maintain identical form to those in Kerr spacetime, i.e.,
\begin{eqnarray}
    \Sigma\dot{t}&=&\frac{1}{\Delta}\left(\tilde{\Sigma}\mathcal{E}-2 M G(r)ar\mathcal{L}\right), \label{eq-geomotion1}\\
    \Sigma\dot{r}&=&\pm\sqrt{\mathcal{R}(r)}, \label{eq-geomotion2}\\
    \Sigma\dot{\theta}&=&\pm\sqrt{\mathcal{Q}(\theta)}, \label{eq-geomotion3}\\
    \Sigma\dot{\phi}&=&\frac{1}{\Delta}\left[2MG(r)ar\mathcal{E}+(\Sigma-2MrG(r))\mathcal{L}\csc^{2}{\theta}\right], \label{eq-geomotion4}
\end{eqnarray}
where
\begin{eqnarray}
    \tilde{\Sigma}&=&(r^2+a^2)^2-a^2\Delta\sin^{2}{\theta},\\
    \mathcal{R}(r)&=&\left[(r^2+a^2)\mathcal{E}-a\mathcal{L}\right]^2-\Delta\left[\mathscr{C}+(a\mathcal{E}-\mathcal{L})^2\right] \nonumber\\
    &~~&+\mu r^2\Delta, \\
    \mathcal{Q}(\theta)&=&\mathscr{C}+(a^2\mathcal{E}^2-\mathcal{L}^2\csc^{2}\theta)\cos^{2}\theta \nonumber\\
    &~~&+\mu a^{2}\cos^{2}{\theta},
\end{eqnarray}
$\mathcal{E},\,\mathcal{L}$ are the conserved energy and orbital angular momentum per unit mass of the particle, due to the stationary and axially symmetric properties of the RGI-Kerr spacetime. $\mathscr{C}$ is the Carter constant due to the separability of the Hamilton-Jacobi equation of the action. $\mu=-1,\,0\,,1$ gives timelike, null and spacelike geodesics, respectively. We consider only $\mu=-1$.

\subsection{The particle motions on the equatorial plane}

To start, we begin with the investigations of the particle motions on the equatorial plane, i.e., $\theta=\pi/2,\,$ and $\dot{\theta}=0$. Then, the equation of radial motion can be expressed as
\begin{eqnarray}\label{eq:ramo2}
    \dot{r}^{2}&=&\mathcal{E}^{2}-\frac{\Delta r-r(a^2\mathcal{E}^2-\mathcal{L}^2)-2MG(r)(a\mathcal{E}-\mathcal{L})^2}{r^{3}}\nonumber \\
    &=&\mathcal{E}^{2}-V_{eff}(r).
\end{eqnarray}
We now focus on two special bound orbits, i.e., the marginally bound orbits (MBO) which specifies the finally captured unstable circular orbit around the black hole formed by a
timelike particle falling freely from the infinity with $\dot{r}=0,\,\mathcal{E}=1$, and the innermost stable orbits (ISCO) which is the closest stable circular orbits that a particle can orbit the black hole.

\begin{figure*}
    \centering
    \includegraphics[width=0.8\textwidth]{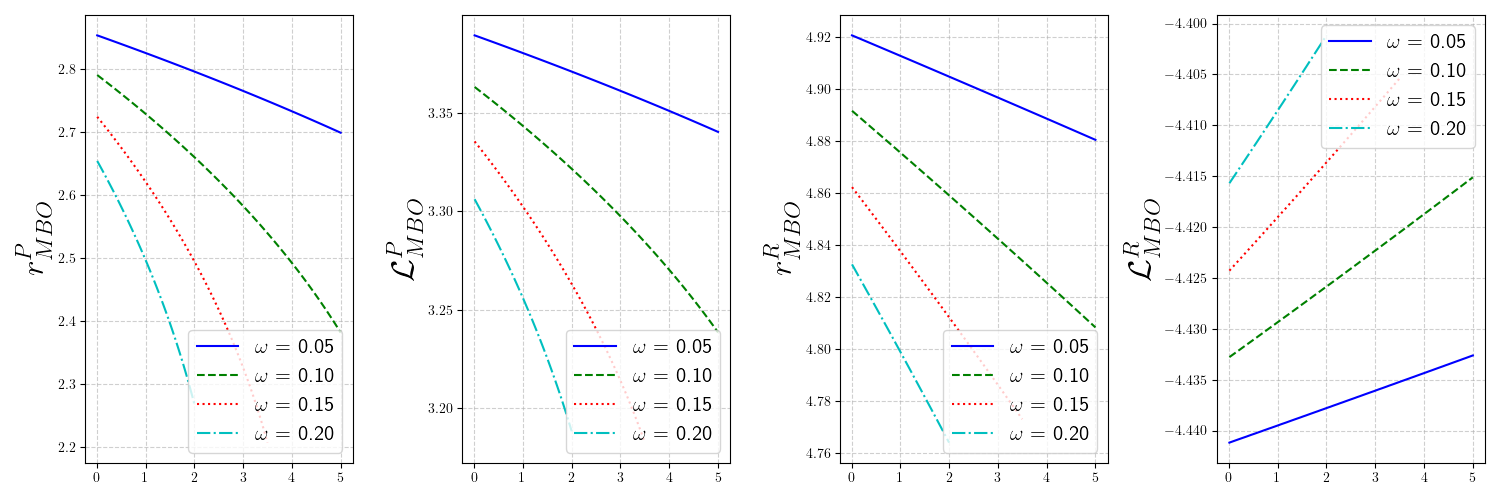}
    \caption{MBOs $r_{MBO}$ and the corresponded angular momentum $\mathcal{L}_{MBO}$ as functions of $\gamma$ when $a=0.5$. The right superscripts $P$ and $R$ represent the prograde and retrograde orbits, respectively.}
    \label{fig:mbo}
\end{figure*}

For MBO, the determined conditions are 
\begin{eqnarray}
    V_{eff}(r)=1,\,V_{eff}'(r)=0,
\end{eqnarray}
where the prime represents the derivative against $r$. In Fig. \ref{fig:mbo} we show the effects of $\gamma$ and $\omega$ on the MBOs $r_{MBO}$ and the critical angular momentum $\mathcal{L}_{MBO}$ focusing on $a=0.5$, where the MBOs of the timelike particles are either prograde or retrograde relative to the black hole's spin direction. Clearly, such behaviors are similar to the outer horizon with the influences of varying $\gamma$ and $\omega$.

\begin{figure*}
    \centering
    \includegraphics[width=0.8\textwidth]{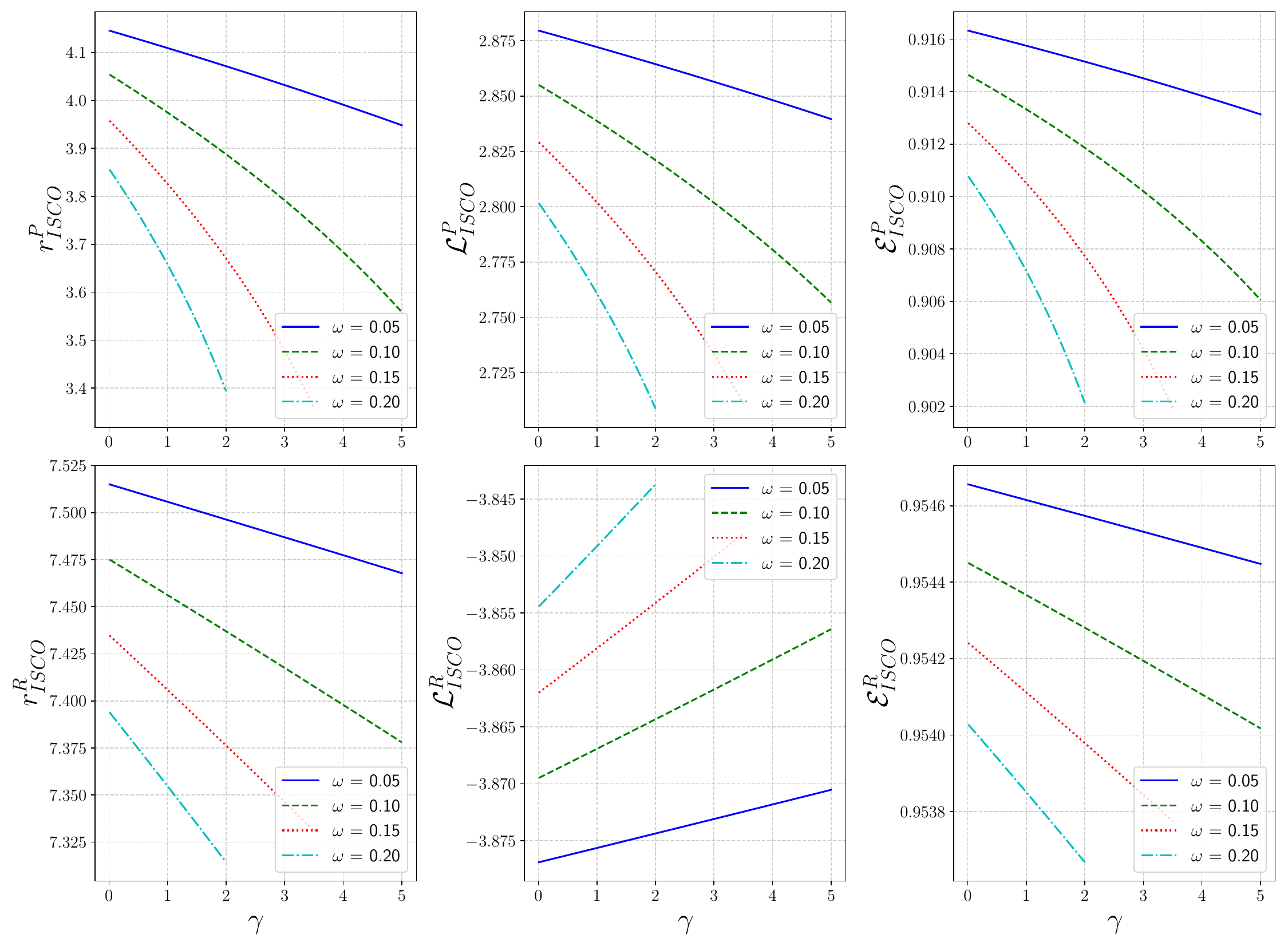}
    \caption{$r_{ISCO}$, $\mathcal{E}_{ISCO}$ and $\mathcal{L}_{ISCO}$ as functions of $\gamma$. The right superscripts $P$ and $R$ represent the prograde and retrograde orbits, respectively. Here $a=0.5$ but for other spin parameters one can find similar behaviors.}
    \label{fig:isco}
\end{figure*}

For ISCO, the determined conditions are now 
\begin{eqnarray}
    V_{eff}(r)=\mathcal{E}^2,\,V_{eff}'(r)=0,\,V_{eff}''(r)=0.
\end{eqnarray}
As expected, $r_{ISCO}$, $\mathcal{E}_{ISCO}$ and $\mathcal{L}_{ISCO}$ all exhibit  analogous dependence on both $\omega$ and $\gamma$ as $r_{MBO}$, see Fig.{\ref{fig:isco}} where we plot the behaviors of $r_{ISCO}$, $\mathcal{E}_{ISCO}$ and $\mathcal{L}_{ISCO}$ as functions of $\gamma$.

In addition to the two aforementioned special types of bound orbits, there exist other distinctive bound orbits, namely precessing and periodic orbits. For the timelike particles moving on the equatorial plane, the motions are completely determined by the $r$ and $\phi$ motions, and can be simply described by an unique number $q$ as \cite{Levin:2008mq}
\begin{eqnarray}\label{eq:qdefin}
    q=\frac{\Delta\phi}{2\pi}-1,
\end{eqnarray}
where $\Delta\phi$ is the accumulated azimuth between two turning points $r_{1},\,r_{2}$ of the bound orbit during a radial period
\begin{eqnarray}\label{eq:azimuth}
    \Delta\phi=\oint d\phi=2\int_{r_2}^{r_1}\frac{d\phi}{dr}dr.
\end{eqnarray}
Apparently, when $q$ is a rational number the timelike particle will move in  periodic orbits and return to its initial location exactly after a finite time. Otherwise, the timelike particle will run a precessing orbit around the black hole and the precession per revolution  can be expressed as $\Delta\chi=\Delta\phi-2\pi$. 

The turning points $r_1$ and $r_2$ are the two roots of $\dot{r}^2=0$, where the energy and the angular momentum of the timelike particle should satisfy $\mathcal{E}_{ISCO}<\mathcal{E}<\mathcal{E}_{MBO}$, $\vert\mathcal{L}_{ISCO}\vert<\vert\mathcal{L}\vert<\vert\mathcal{L}_{MBO}\vert$. More precisely, if the condition $\dot{r}^2=0$ is required to admit at least two distinct solutions, then the allowable range of either $\mathcal{E}$ (when fixing $\mathcal{L}$) or $\mathcal{L}$ (when fixing $\mathcal{E}$) will necessarily be narrower than the original (ISCO, MBO) parameter region. Since our primary objective is to investigate the influence of $\omega$ and $\gamma$ on the trajectories of timelike particles, in the subsequent numerical calculations of this section we will focus on two specific scenarios, i.e., we fix $\mathcal{L}$ while varying $\mathcal{E}$, or we maintain $\mathcal{E}$ while varying $\mathcal{L}$. 

\begin{figure*}
    \centering
    \includegraphics[width=0.8\linewidth]{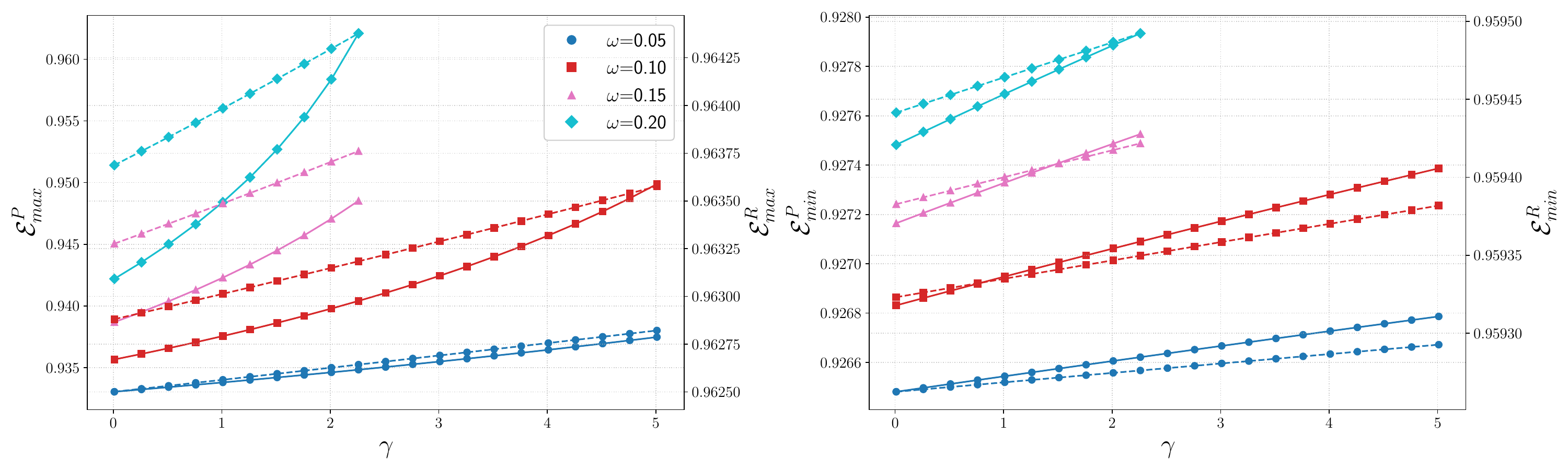}
    \caption{The energy $\mathcal{E}_{m}$ as functions of $\gamma$ with fixed $\mathcal{L}_{m}=3$ (prograde) or $\mathcal{L}_{m}=-4$ (retrograde), where the spin parameter $a=0.5$. The solid lines correspond to prograde motions (right superscripts $P$) while the dashed lines represent retrograde motions (right superscripts $R$).}
    \label{fig:emaxvsgamma}
\end{figure*}

\begin{figure*}
    \centering
    \includegraphics[width=0.8\linewidth]{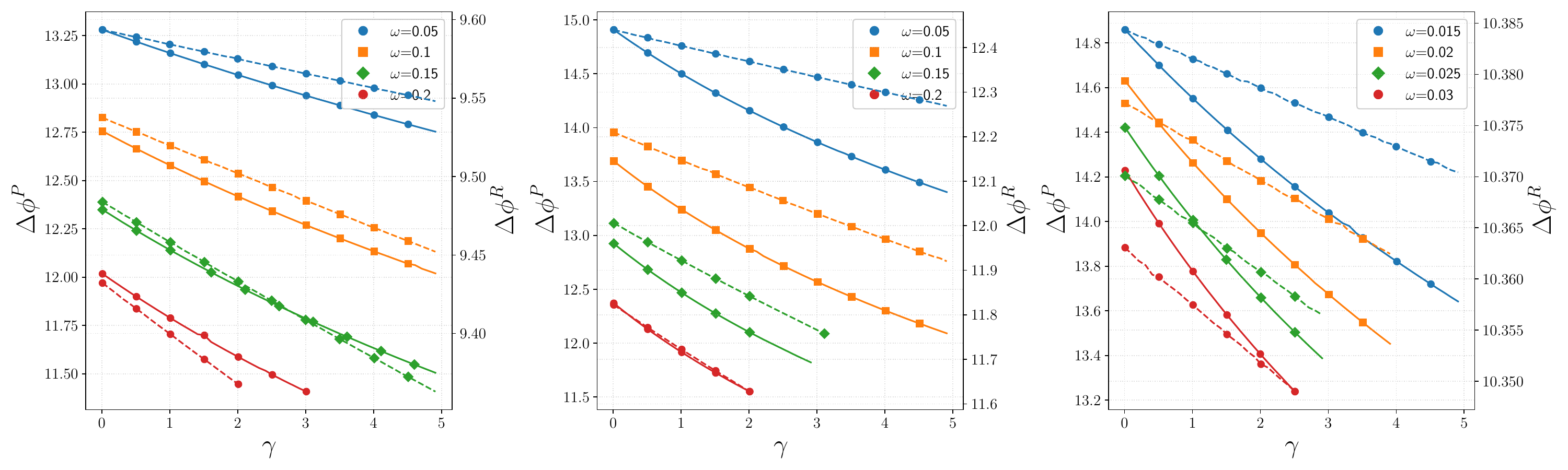}
    \caption{The accumulated azimuth $\Delta\phi$ as a function of $\gamma$ for different selected parameters $(a,\, \omega,\,\mathcal{L},\,\mathcal{E})$. From left to right, the parameter sets are selected as: (Left) $(a=0.1,\,\mathcal{L}^{P}_{m}=3.5,\,\mathcal{E}^{P}_{m}=0.95,\,\mathcal{L}^{R}_{m}=-4,\,\mathcal{E}^{P}_{m}=0.97)$; (Middle) $(a=0.5,\,\mathcal{L}^{P}_{m}=3,\,\mathcal{E}^{P}_{m}=0.93,\,\mathcal{L}^{R}_{m}=-4,\,\mathcal{E}^{P}_{m}=0.96)$; (Middle) $(a=0.9,\,\mathcal{L}^{P}_{m}=2.35,\,\mathcal{E}^{P}_{m}=0.90,\,\mathcal{L}^{R}_{m}=-4.5,\,\mathcal{E}^{P}_{m}=0.975)$. The solid lines correspond to prograde motions (right superscripts $P$) while the dashed lines represent retrograde motions (right superscripts $R$).}
    \label{fig:phivsgamma}
\end{figure*}

To examine the influence of $\omega$ and $\gamma$ on $\Delta\phi$ for precessing orbits, we fix $\mathcal{L}_{m}$ and select a corresponding $\mathcal{E}_{m}$ such that $\dot{r}^{2}=0$ admits at least two roots across the entire parameter range ($\omega, \gamma$) under consideration. For example, in Fig. \ref{fig:emaxvsgamma} we show the maximum and minimum $\mathcal{E}_{m}$ as functions of $\gamma$  when $\mathcal{L}$ is chosen as $\mathcal{L}_{m}=3$ (prograde) or $\mathcal{L}_{m}=-4$ (retrograde) for different $\omega$ with $a=0.5$. Therefore, selecting $\mathcal{E}_{m}=0.93$ (prograde) or $\mathcal{E}_{m}=0.96$ (retrograde) suffices to guarantee that the condition $\dot{r}^{2}=0$ possesses three distinct roots for all selected $\omega$, in which the smallest root corresponds to the second kind geodesic falling to the horizon, and the intermediate and largest roots represent the turning points of the first kind bound orbits, i.e., the periastron and apoastron, respectively \cite{Chandrasekhar:1985kt}.

\begin{figure*}
    \centering
    \includegraphics[width=0.8\linewidth]{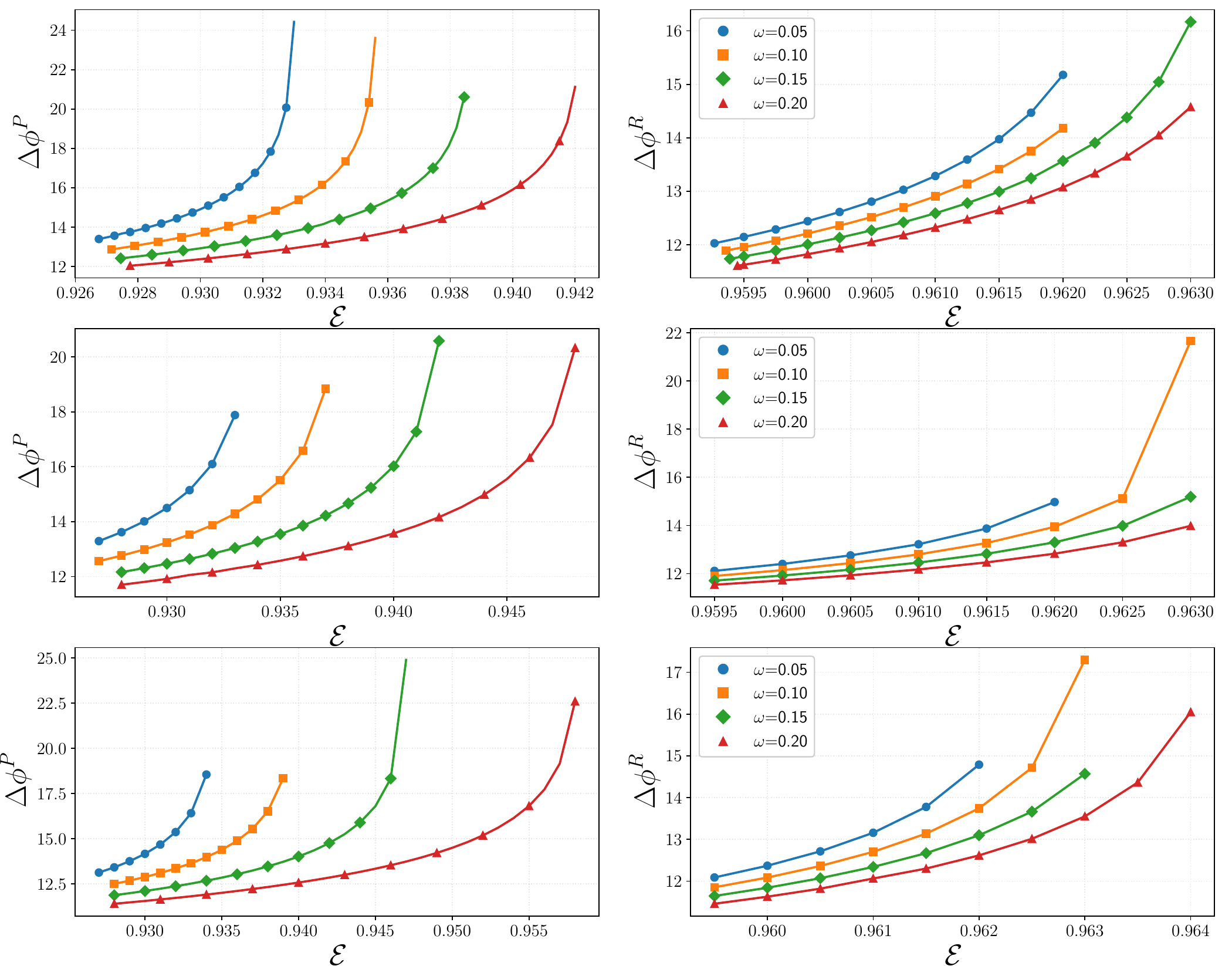}
    \caption{The accumulated azimuth $\Delta\phi$ as a function of $\mathcal{E}$ for different selected parameters $(a=0.5,\, \omega)$ and $\mathcal{L}_{m}=3$ (left), $\mathcal{L}_{m}=-4$ (right). From top to bottom, $\gamma=0.01,\, 1.01,\, 2.01$. The left column correspond to prograde motions while the right are retrograde motions. Note here $\mathcal{E}\in(\mathcal{E}_{min},\mathcal{E}_{max})$, where $\mathcal{E}_{max}$ and $\mathcal{E}_{min}$ are shown in Fig. \ref{fig:emaxvsgamma}.}
    \label{fig:deltaphivsEa05}
\end{figure*}

In Fig. \ref{fig:phivsgamma} we present the accumulated azimuth $\Delta\phi$ of the processing orbits as a function of $\gamma$ for different selected parameters $(a,\, \omega,\,\mathcal{L},\,\mathcal{E})$. Unfortunately, regardless of whether the spin parameter $a$ remains unchanged, there is no identical parameter set $(\pm\mathcal{L},\,\mathcal{E})$ that would allow us to focus more on the influence of $\omega$ and $\gamma$ on $\Delta\phi$. Even under the same spin parameter $a$, prograde or retrograde motion will cause $\pm\mathcal{L}$ and $\mathcal{E}$ to change. Therefore, we can only discuss the effect of $\omega$ and $\gamma$ on $\Delta\phi$ under the same spin parameter $a$ and the same particle motion. Nevertheless, a consistent trend can be clearly observed from the figures: irrespective of whether the orbit is prograde or retrograde, and regardless of whether the parameters $\omega$ and $\gamma$ increase either individually or concurrently, the accumulated azimuth $\Delta\phi$ invariably exhibits a monotonic decrease. As previously discussed, the outer horizon radius of RGI-Kerr black holes decreases with increasing parameters $\omega$ and $\gamma$, corresponding to a reduction in the black hole's effective mass. Consequently, this leads to a smaller precession angle for timelike particles.

We will close this subsection by seeking the periodic timelike orbits, where the ratio $q$ has to be a rational number. To achieve this objective, similar to the previous calculation of the precession angle, we can first utilize Eq.(\ref{eq:ramo2}) to determine the period of $r(\phi)$ at different energy levels under fixed $(\mathcal{L},\,a,\,\gamma,\,\omega)$, thereby establishing the relationship between the angle $\phi$ and the energy $\mathcal{E}$. Then by treating $q$ as a function of energy $\mathcal{E}$ and applying definition (\ref{eq:qdefin}), we can then identify the energy values corresponding to rational $q$, which in turn allows us to derive the periodic orbits. Fig. \ref{fig:deltaphivsEa05} present the accumulated azimuth $\Delta\phi$ as a function of $\mathcal{E}$ for different selected parameters $(a=0.5,\, \omega,\,\mathcal{L})$ and $\gamma$. Fig. \ref{fig:phivsgamma} have illustrated that with fixed $a,\,\mathcal{L}$ and $\mathcal{E}$ the azimuth $\Delta\phi$ becomes smaller with increasing $\omega$ or/and $\gamma$, while Fig. \ref{fig:deltaphivsEa05} then illustrates that for given $\gamma$ (or $\omega$) and $\mathcal{L}$, a larger $\omega$ (or $\gamma$) requires a higher energy $\mathcal{E}$ to achieve the same angular $\Delta\phi$. The information presented in these two figures is consistent. Furthermore, due to the limited region of $\mathcal{E}$ for fixed $\mathcal{L}$, not all rational $q$ are allowed. 

\subsection{The general cases}

\begin{figure*}
    \centering
    \includegraphics[width=0.4\linewidth]{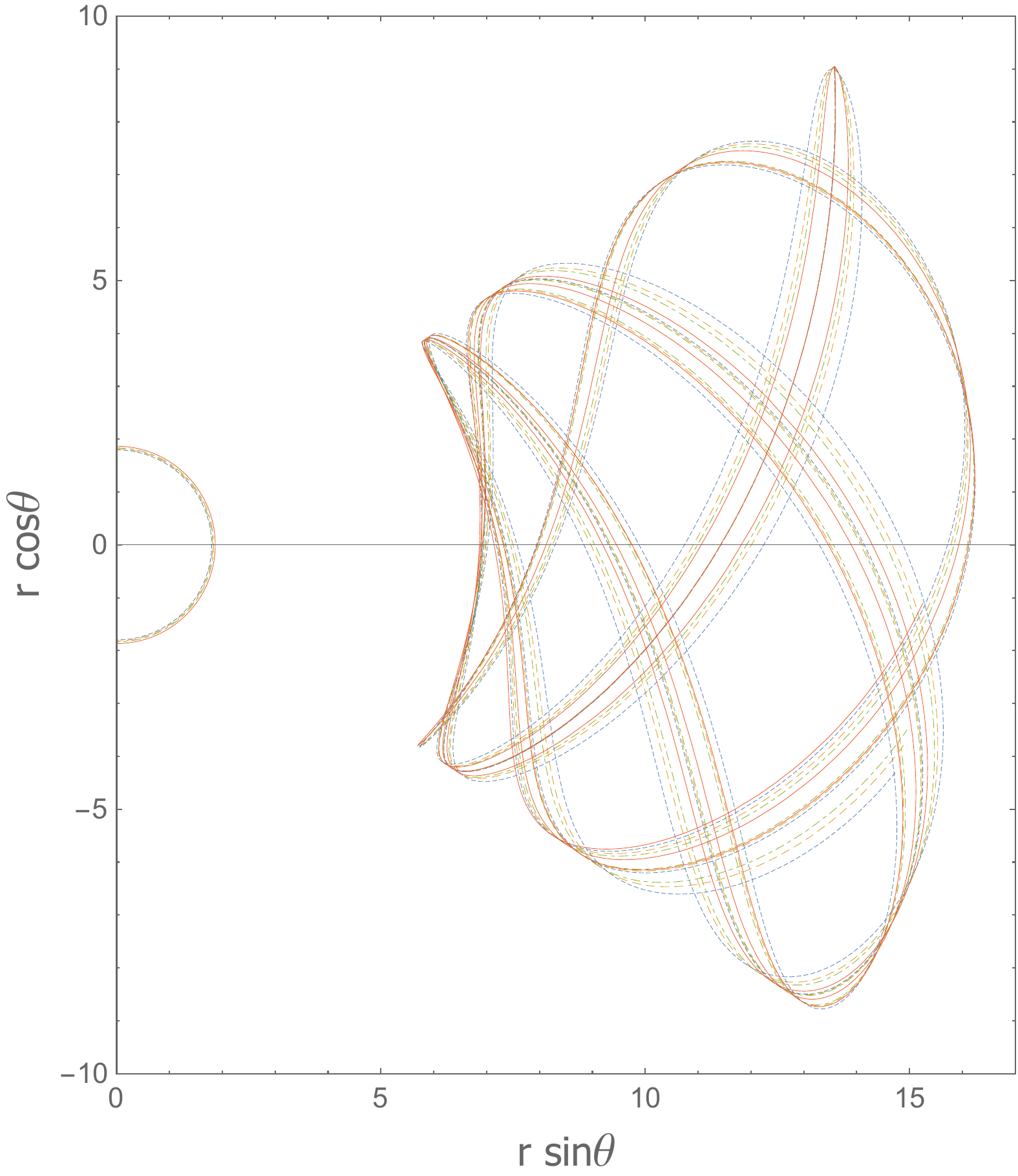}\hspace{0.5cm}
    \includegraphics[width=0.4\linewidth]{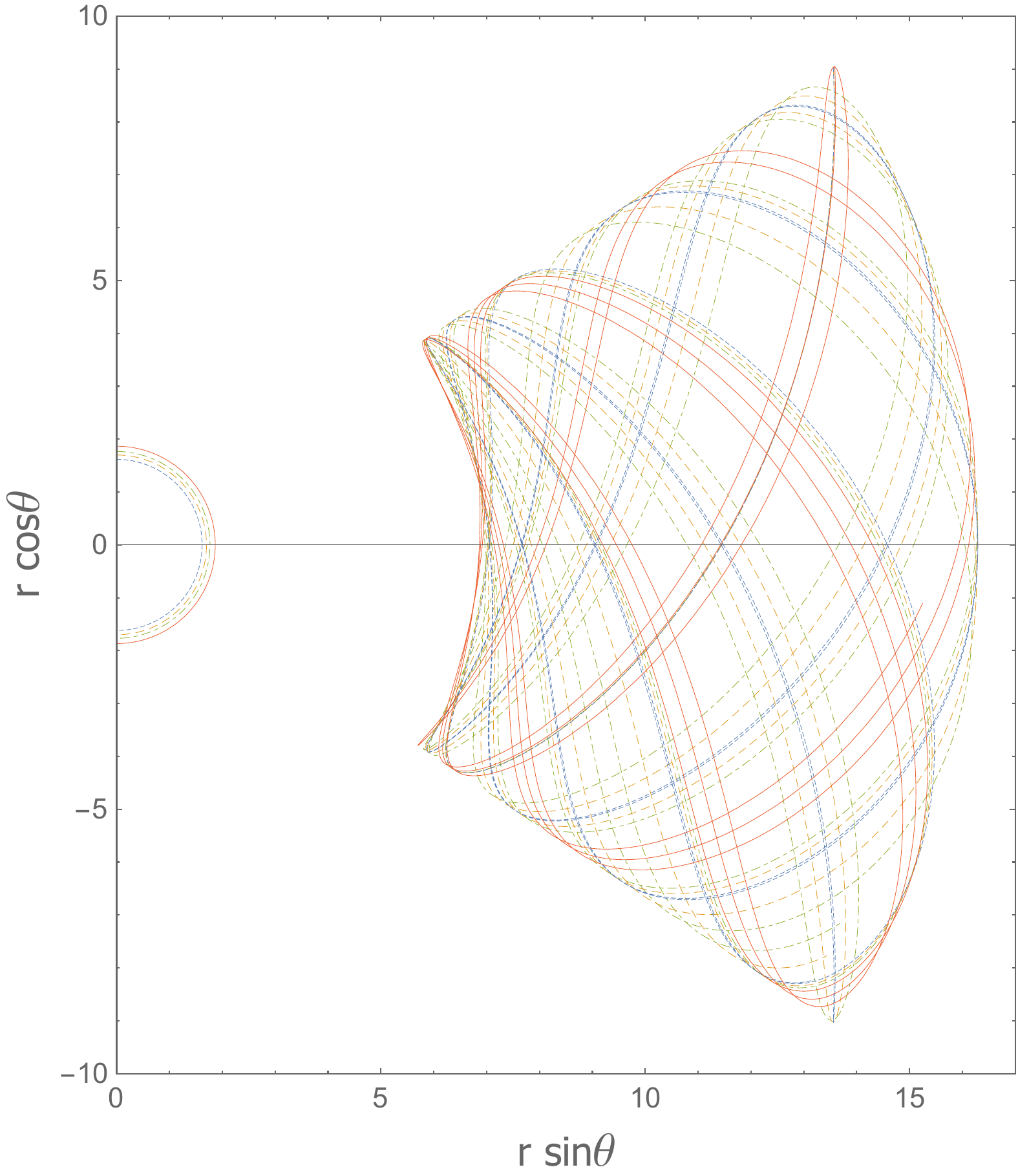}
    \caption{The projections of orbits on the meridian plane for selected parameters $a=0.5,\,\mathcal{E}_{m}=0.96,\,\mathcal{L}_{m}=3, \,\mathscr{C}=4$. Left panel: $\omega=0.05$. Right panel: $\omega=0.15$. In each panel, the dotted, dashed and dot-dashed line correspond to $\gamma=2.01,\,1.01,\,0.01$, respectively. The solid line indicates the orbit for a classical Kerr black hole with $a=0.5,\,\mathcal{E}_{m}=0.96,\,\mathcal{L}_{m}=3, \,\mathscr{C}=4$. The half circles correspond to the horizon.}
    \label{fig:noneqa05}
\end{figure*}

We now consider the general bound orbits of the timelike particles outside the equatorial plane. However, we will provide only a brief discussion of such orbits, leaving a more detailed exploration for future work. Due to the stationary and axisymmetric nature of the RGI-Kerr black hole, along with the fact that the coordinate functions $t(\tau)$ and $\phi(\tau)$ in the geodesic equations (\ref{eq-geomotion1},\ref{eq-geomotion2},\ref{eq-geomotion3},\ref{eq-geomotion4}) can be fully expressed in terms of $r(\tau)$ and $\theta(\tau)$, we can restrict our discussion of such general timelike geodesics to the meridian plane generated by $r$ and $\theta$.

To numerically investigate the influences of $\omega$ and $\gamma$ on these general geodesics, we fix the parameter values $a=0.5,\,\mathcal{E}_{m}=0.96,\,\mathcal{L}_{m}=3$ (for prograde case), $\mathscr{C}=4$ and $\omega=(0.05,\,0.15), \,\gamma=(0.01,\,1.01,\,2.01)$. In fact, for other pre-fixed parameters, there exists a minimum energy $\mathcal{E}_{m}$ below which no bound orbits can exist. In Fig. \ref{fig:noneqa05} we have shown the projections of orbits on the meridian plane for our selected parameters. As can be easily observed from the figure, for small values of $\omega$, the deviation from the classical Kerr black hole induced by $\gamma$ is less pronounced than that observed under larger $\omega$ values. This characteristic is also reflected in the gravitational waveforms emitted by timelike particles on periodic orbits under different parameter configurations. Additionally, these projected trajectories indicate that the corresponding timelike particle orbits are not periodic.

\section{THE GRAVITATIONAL WAVE FROM PERIODIC ORBITS}\label{sec:GW}

In this section we move on to provide a preliminary exploration of the gravitational radiation emitted by the periodic orbits of a test particle orbiting a supermassive RGI-Kerr black hole, by assuming that the smaller object has a mass extremely smaller than the central black hole and moves on the equatorial plan. As aforementioned such EMRI observations are able to detect the fundamental physics for black holes in a large class of theories \cite{Maselli:2021men,Fell:2023mtf,Zi:2024lmt}. Concurrently, these EMRI waveforms are primarily generated through three distinct methodologies, i.e., the solutions of the Teukolsky equation derived from black hole perturbation theory, the self-consistent post-Newtonian analytic approaches and the so-called kludge approximation schemes. Anyway, to implement these methods in the present study, we must first examine the corresponding action. The effective average action in the Einstein-Hilbert truncation leads to the improved Einstein's equations (IEE) reading
\begin{eqnarray}
    G_{\mu\nu}=8\pi G_{k} T_{\mu\nu}-\Lambda_{k}g_{\mu\nu}+\alpha\Delta T^{RG}_{\mu\nu},
\end{eqnarray}
where $G_{k},\,\Lambda_{k}$ stand for the running coupling constants and 
\begin{eqnarray}
    \Delta T^{RG}_{\mu\nu}=G_{k}(\nabla_{\mu}\nabla_{\nu}-g_{\mu\nu}\Box)G_{k}^{-1}.
\end{eqnarray}
$\alpha =1$ indicates that the RG improvement is performed at the level of the action, whereas $\alpha=0$ if it is employed at the level of field equations \cite{Reuter:2003ca,Reuter:2004nv,Platania:2019qvo,Platania2023}. Clearly, the metric (\ref{eq:metric}) does not satisfy this equation. In fact, this metric can be regarded as a solution to Einstein's equations with a pseudo-matter field \cite{Reuter:2010xb}
\begin{eqnarray}\label{eq:effectiveEE}
    G_{\mu\nu}=8\pi G_0 \tilde{T}_{\mu\nu},
\end{eqnarray}
where the components of $\tilde{T}_{\mu\nu}$ have also been displayed in Ref. \cite{Reuter:2010xb}. Therefore, in a simplified interpretation, the gravitational effects induced by $G(r)$ can be equivalently treated as those generated by an additional pseudo-matter field \cite{Reuter:2003ca}. Considering that the metric (\ref{eq:metric}) reduces to the Kerr spacetime in the far-field approximation $(r\rightarrow{\infty})$, we can thus employ the numerical kludge method to approximately derive the gravitational waveforms generated by the secondary compact object moving along the orbits under the EMRI assumption.

{However, given that our analysis is confined to bound orbits and primarily examines the influence of the parameters $\gamma$ and $\omega$ on orbital dynamics, we have neglected the loss of the gravitational energy radiated by the orbital motion. In practice, this corresponds to retaining only the leading-order term within the post-Newtonian approximation. Following Refs. (\cite{Babak:2006uv,Poisson:2014kt,Maselli:2021men,Liang:2022gdk}), the gravitational wave radiated from these orbits can be computed using the following  quadratic order formula:}
\begin{eqnarray}\label{GWfirst}
    h_{ij}=\frac{2G_0}{c^4D_{L}}\ddot{I}^{ij},
\end{eqnarray}
{where $I_{ij}=\int d^{3}xT^{tt}(t,x^i)x^ix^j$ is the mass quadrupole moment given in terms of the source stress-energy tensor $T^{tt}(t,x^i)=m\delta^{3}(x^i-z^{i}(t))$. Note here $x^i$ are Cartesian spatial coordinates as defined in Ref. (\cite{Babak:2006uv}) and $z^i(t)$ is the worldline of the secondary object. For a binary EMRI system of orbiting bodies, Eq. (\ref{GWfirst}) becomes 
\begin{eqnarray}
    h_{ij}=\frac{4G_{0}\eta (M+m)}{c^{4}D_{L}}\left(v_{i}v_{j}-\frac{G_{0}(M+m)}{r}n_{i}n_{j}\right),
\end{eqnarray}
where $\eta=Mm/(M+m)^{2}$ is the symmetric mass ratio, $M$ and $m$ are the masses of the RGI-Kerr black hole and the secondary object, respectively. $D_{L}$ is the luminosity distance of the EMRI system, $\boldsymbol{v}$ is the spatial relative velocity of the secondary object and $\boldsymbol{n}$ is the unit vector pointing to the radial direction associated to the motion of the secondary object \cite{Poisson:2014kt}. }

{Now to describe the orbital motion one can define an orbital eccentricity $e$ and semi-latus rectum $p$ using the conventional Keplerian definitions
\begin{eqnarray}
    r_p=\frac{p}{1+e},\qquad r_{a}=\frac{p}{1-e},
\end{eqnarray}
and then work in terms of a new angular variable $\chi$ as
\begin{eqnarray}
    r=\frac{p}{1+e\cos{\chi}}.
\end{eqnarray}
Clearly, as the parameter $\chi$ transforms between 0 and $2\pi$, the coordinate $r$ travels back and forth between $r_{a}$ and $r_p$. In the orbital plane, by introducing a ``orbit-adapted" system $(x,y,z)$ where the origin is the barycenter, $(x-y)$ plane coincides with the orbital plane and the $z$-axis points in the direction of the angular momentum vector, $x$-axis points to the apoastron,  one will have the unit vectors
\begin{eqnarray}
  \bm{n}=\left(\cos{\chi},\,\sin{\chi},\,0\right),\quad \bm{\lambda}=\left(-\sin{\chi}\,,\cos{\chi},\,0\right)
\end{eqnarray}
and then
\begin{eqnarray}
    \bm{r}=r\bm{n},\qquad \bm{v}=\dot{r}\bm{n}+r\dot{\chi} \bm{\lambda}.
\end{eqnarray}
}

\begin{figure*}
    \centering
    \includegraphics[width=0.8\linewidth]{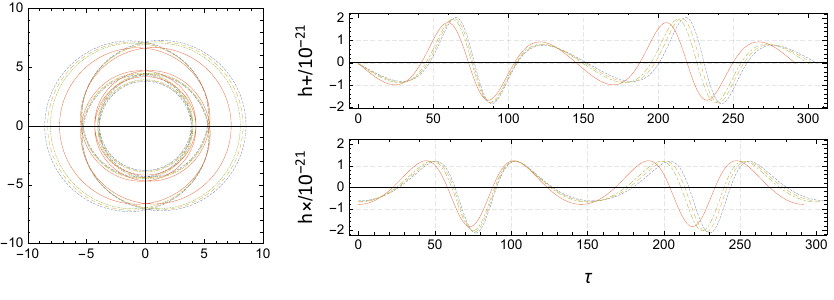}
    \includegraphics[width=0.8\linewidth]{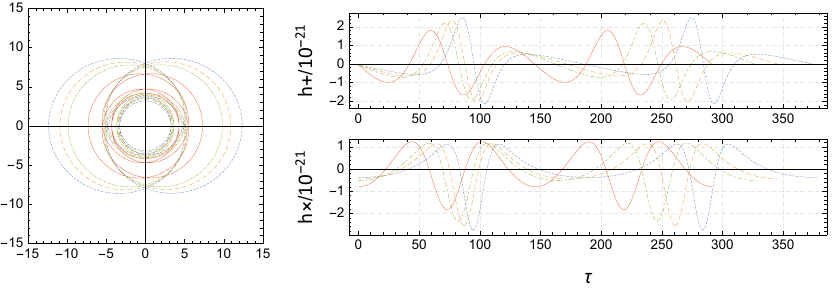}
    \caption{The prograde periodic bound orbits and their gravitational waveforms with $q=3/2,\,a=0.5,\,\mathcal{L}_{m}=3$.  From top to down, $\omega=0.05,\,0.15$. The dotted, dashed and dot-dashed line correspond to $\gamma=2.01,\,1.01,\,0.01$, respectively. The solid line indicates the prograde periodic bound orbits and the gravitational waveforms for a classical Kerr black hole ($q=3/2,\,a=0.5$).}.
    \label{fig:periodqa05}
\end{figure*}

\begin{figure*}
    \centering
    \includegraphics[width=0.8\linewidth]{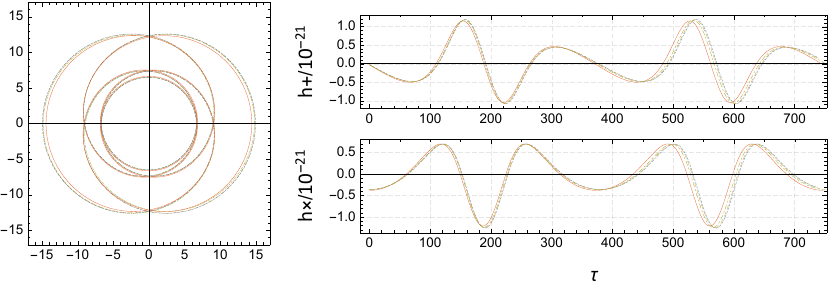}
    \includegraphics[width=0.8\linewidth]{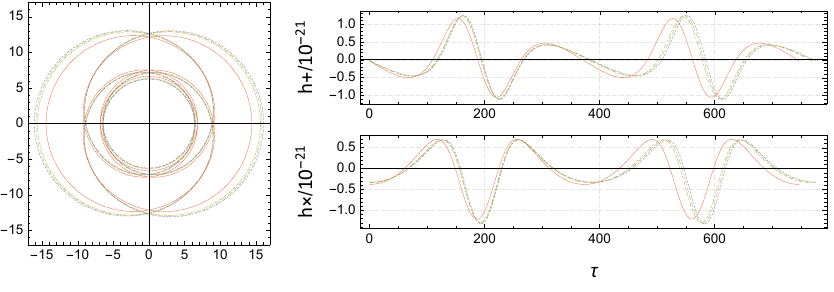}
    \caption{The retrograde periodic bound orbits and their gravitational waveforms with $q=3/2,\,a=0.5,\,\mathcal{L}_{m}=-4$. From top to down, $\omega=0.05,\,0.15$. The dotted, dashed and dot-dashed line correspond to $\gamma=2.01,\,1.01,\,0.01$, respectively. The solid line indicates the retrograde periodic bound orbits and the gravitational waveforms for a classical Kerr black hole ($q=3/2,\,a=0.5$).}.
    \label{fig:periodqa05r}
\end{figure*}

Based on the orbit adapted coordinates, one can further introduce a ``detector-adapte'' Cartesian coordinate system $(X, Y, Z)$ as
\begin{eqnarray}
    \boldsymbol{e}_{X}&=&(\cos{\zeta},\,-\sin{\zeta},\,0),\\
    \boldsymbol{e}_{Y}&=&(\cos{\iota}\sin{\zeta},\,\cos{\iota}\cos{\zeta},\,-\sin{\iota}),\\
    \boldsymbol{e}_{Z}&=&(\sin{\iota}\sin{\zeta},\,\sin{\iota}\cos{\zeta},\,\cos{\iota}),
\end{eqnarray}
{where $\iota$ denotes the inclination angle between the EMRI’s orbital angular momentum and the line of sight, and $\zeta$ denotes the latitudinal angle between the pericenter and the line of nodes, as measured in the orbital plane. Such system should have the same origin with the orbit system $(x, y, z)$, and the $Z$-axis points in the direction of the gravitational wave detector. The $(X-Y)$ plane is orthogonal to the $Z$-axis and coincides with the plane of the sky from the detector’s point of view, and the $X$-axis points toward the ascending node, the point at which the orbit cuts the plane from below \cite{Poisson:2014kt}. Then by adopting $\bm{e}_{X}$ and $\bm{e}_{Y}$ as the vectorial basis in the transverse subspace, the gravitational wave polarizations, $h_{+}$ and $h_{\times}$, take the forms}
\begin{eqnarray}\label{GWpolar}
    h_{+}&=&\frac{1}{2}\left(e_{X}^{j}e_{X}^{k}-e_{Y}^{j}e_{Y}^{k}\right)h_{jk},\\
    h_{\times}&=&\frac{1}{2}\left(e_{X}^{j}e_{Y}^{k}+e_{Y}^{j}e_{X}^{k}\right)h_{jk}.
\end{eqnarray}
{Inserting the expressions of $\bm{n}$ and $\bm{\lambda}$ in the detector-adapted coordinates
\begin{eqnarray}
\bm{n}&=&\left[\cos{ \tilde{\zeta}},\,\cos{\iota}\sin{ \tilde{\zeta}}\,,\sin{\iota}\sin{ \tilde{\zeta}}\right],\\
    \bm{\lambda}&=&\left[-\sin{ \tilde{\zeta}},\,\cos{\iota}\cos{ \tilde{\zeta}}\,,\sin{\iota}\cos{ \tilde{\zeta}}\right],\\
    \tilde{\zeta}&=&\zeta+\chi,
\end{eqnarray}
within Eq. (\ref{GWpolar}) one will have
\begin{eqnarray}
    h_{+}&=&-h_0(1+\cos^{2}{\iota})\left[\cos{(2\chi+2\zeta)+\frac{5}{4}e\cos{}(\chi+2\zeta)}\right.\nonumber\\
    &\quad&+\left.\frac{1}{4}e\cos{(3\chi+2\zeta)}+\frac{1}{2}e^2\cos{2\zeta}\right] \nonumber\\
    &\quad&+\frac{1}{2}e\sin^{2}{\iota}(\cos{\chi}+e),  \\
    h_{\times}&=&-2h_0\cos{\iota}\left[\sin{(2\chi+2\zeta)}+\frac{5}{4}e\sin{(\chi+2\zeta)}\right.\nonumber \\
    &\quad&+\left.\frac{1}{4}e\sin{(3\chi+2\zeta)}+\frac{1}{2}e^2\sin{2\zeta}\right],\\
    h_0&=&\frac{2G_0}{c^2D_{L}p}\frac{m}{M}.
\end{eqnarray}}

To analyze the gravitational waveform of different periodic orbits and the influences of $\omega,\,\gamma$ on the waveforms we consider an EMRI system that consists of a supermassive RGI-Kerr black hole with mass $M=10^7M_{\odot}$ and a secondary object with mass $m=10M_{\odot}$, and adopt the leading order post-Newtonian approximation. For other parameters we set $D_{L}=200 \,\text{Mpc},\,\iota=\zeta=\pi/4$. Moreover, since $\omega$ and $\gamma$ show mainly monotonic effects, we restrict parameters to $\omega=(0, 0.05, 0.15),\,  \gamma=(0.01, 1.01, 2.01)$, with periodic orbit parameter $q= 3/2$ for selected spin parameter $a=0.5$. {Note that the orbit parameter $q$ defined in \eqref{eq:qdefin} describes  the ratio between the oscillations frequencies in the radial and azimuthal motions, which can be specified by three integers denoting the zoom number, whirl number and the vertex number of the orbit \cite{Levin:2008mq}. In a fixed background, the periodic orbit represented by $q$ is determined by the energy $\mathcal{E}$ and orbital angular momentum $\mathcal{L}$ of the particle.}

In Figs. \ref{fig:periodqa05} and \ref{fig:periodqa05r} we present the influence of different values of $\omega$ and $\gamma$ on the periodic orbits of the secondary object and the corresponding gravitational waveforms for a spin parameter of $a=0.5$ and an orbital parameter $q=3/2$. It can be observed that in the prograde case, as $\omega$ decreases, both the periodic orbits and the resulting gravitational waveforms approach those of the classical Kerr black hole (the solid line). Moreover, larger values of $\gamma$ lead to more pronounced deviations from the Kerr model. In contrast, for retrograde orbits, the effects of $\omega$ and $\gamma$ on both the orbital structure and the waveforms are considerably less significant. These findings are consistent with the behavior illustrated in the right column of Fig. \ref{fig:deltaphivsEa05}, in which for a fixed angular momentum $\mathcal{L}$, the energy $\mathcal{E}$ required to sustain the $q=3/2$ periodic orbit exhibits minimal variation across different $\omega$ and $\gamma$ values, suggesting only minor modifications in the orbital characteristics—a conclusion corroborated by the results shown in Fig. \ref{fig:periodqa05r}.

\begin{figure*}
    \centering
    \includegraphics[width=\linewidth]{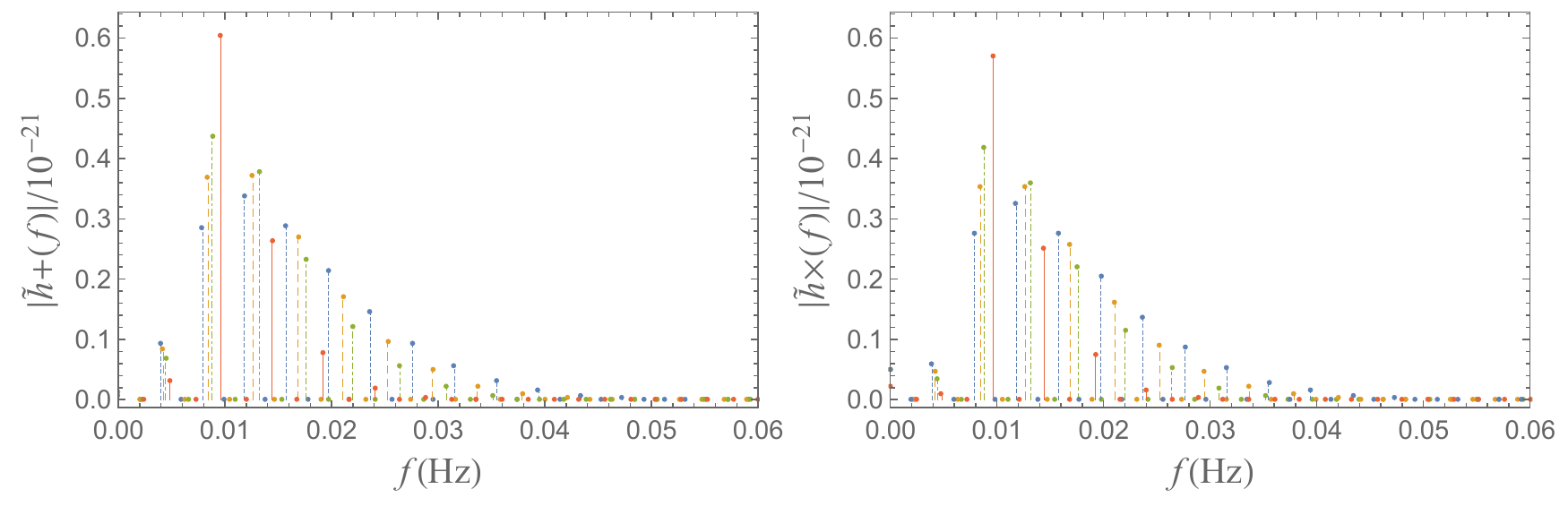}
    \caption{The absolute frequency spectra $|\tilde{h}_{+}(f)|$ and $|\tilde{h}_{\times}(f)|$ corresponding to the gravitational waveforms from Fig. \ref{fig:periodqa05} with $q=3/2,\,a=0.5,\,\mathcal{L}_{m}=3, \, \omega=0.15$. The dotted, dashed and dot-dashed line correspond to $\gamma=2.01,\,1.01,\,0.01$, respectively. The solid line is for a classical Kerr black hole ($q=3/2,\,a=0.5$). }
    \label{fig:frequencya05}
\end{figure*}

The gravitational waveforms can be further analyzed through the corresponding frequency spectra $\tilde{h}_{+}(f)$ and $\tilde{h}_{\times}(f)$ by applying a discrete Fourier transform (DFT) to the time-domain gravitational waveforms $h_{+}$ and $h_{\times}$, respectively. This approach provides a detailed examination of the signal's frequency distribution, revealing how the particle's periodic orbital motion influences the gravitational wave structure, and is also particularly significant for both space-based and ground-based gravitational wave detectors. Considering that the results presented earlier have already demonstrated the influence of $\omega$ and $\gamma$ on the gravitational waveforms, for the sake of brevity, in Fig. \ref{fig:frequencya05} we only show the low-frequency distribution of periodic orbits for different $\gamma$ values at $\omega=0.15$, i.e., the second row from Fig. \ref{fig:periodqa05}. The results remain consistent with the conclusions drawn from Fig. \ref{fig:periodqa05}, namely that when $\omega$ is relatively large, a higher $\gamma$ value leads to a greater deviation from the classical Kerr case. 

Meanwhile, in gravitational wave astronomy, the sensitivity curve, also known as the noise power spectral density curve, is commonly adopted to quantitatively characterize the detection capability of detectors \cite{Huang:2025czx}. This curve specifies the detector's equivalent noise level across specific Fourier frequency ranges, with the vertical axis typically represented in logarithmic coordinates of amplitude spectral density or characteristic strain which is defined by the frequency spectra $\tilde{h}_{+}(f)$ and $\tilde{h}_{\times}(f)$ as \cite{Finn:2000sy}
\begin{eqnarray}
    S_{c}(f)=2f\sqrt{\left(|\tilde{h}_{+}(f)|^{2}+|\tilde{h}_{\times}(f)|^{2}\right)}. \label{eq:strain}
\end{eqnarray}

\begin{figure*}
    \centering
    \includegraphics[width=0.7\linewidth]{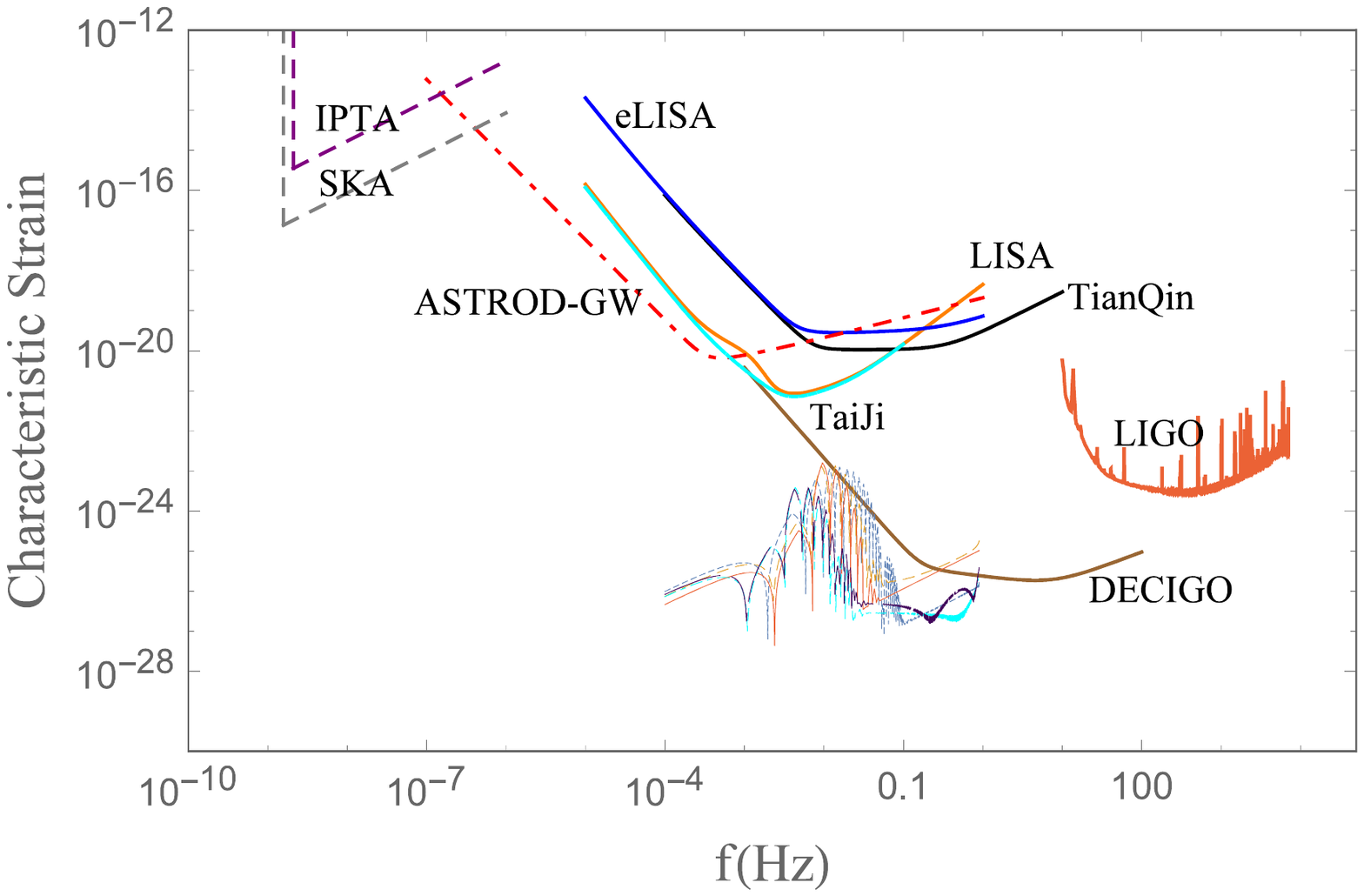}
    \caption{Comparison of the characteristic strain of gravitational waves emitted from the periodic orbits around the RGI-Kerr black hole to the sensitivity curves of various detectors. We fix $a=0.5,\,q=3/2,\,\gamma=2.01$. For prograde case, $\mathcal{L}_{m}=3$ and we set $\omega=0.05$ (dashed yellow line), $\omega=0.15$ (dotted blue line), $\omega=0$ (classical Kerr case, solid red line). For retrograde case, $\mathcal{L}_{m}=-4$ and $\omega=0.05$ (dot-dashed cyan line), $\omega=0.15$ (long dashed purple line).}
    \label{fig:strain}
\end{figure*}

We  then plot the characteristic strain $S_{c}(f)$ of the gravitational waves emitted from timelike test particles around a RGI-Kerr black hole, and compare with the sensitivity curves of various detectors, such as LISA \cite{Robson:2018ifk}, eLISA \cite{Amaro-Seoane:2012vvq}, TianQin \cite{TianQin:2015yph,Torres-Orjuela:2023hfd}, LIGO \cite{Essick:2025zed,LIGOScientific:2025slb,ligo_scientific}, ASTROD-GW \cite{Ni:2012eh}, DECIGO \cite{Ishikawa:2020hlo,Iwaguchi:2020cxa}, TaiJi \cite{Hu:2017mde,Luo:2019zal,Gong:2021gvw,Liu:2023qap}, SKA $\&$ IPTA \cite{Hobbs2010CQGra}, see Fig. \ref{fig:strain}. This visual comparison shows that the characteristic frequency of the periodic orbits $(q=3/2)$ around the RGI-Kerr black hole is concentrated between $10^{-3}$ Hz $\sim$ $0.1$ Hz and the corresponding characteristic strains exceed the
sensitivity curves of DECIGO. However, these frequency ranges still fall within the most sensitive detection bands of observatories such as LISA, eLISA, Taiji and TianQin. Therefore, with further improvements in detector sensitivity, it is anticipated that more gravitational wave detectors in the future may be able to observe the gravitational wave signals emitted by EMRIs in the RGI-Kerr black hole background.

\section{Summaries and Remarks}\label{sec:conclusion}
In this article, we have investigated the bound orbits of the secondary object in the EMRI system and the gravitational waveforms emitted from these orbits in a RGI-Kerr background. In such background, the classical Newtonian gravitational constant is replaced by a scale-dependent running gravitational coupling, and the geometric properties of spacetime are consequently characterized primarily by the parameters $\omega$ and $\gamma$ \cite{Held:2019xde}. Thus the  properties of the timelike geodesics and the gravitational waveforms can also be characterized by these two parameters.

First, we examined the geodesic properties of timelike particles under the RGI-Kerr background, specifically focusing on bound orbits. The results indicate that for both the MBO and the ISCO, their radii decrease as either $\omega$ or $\gamma$ increases. Correspondingly, the associated angular momentum and energy also diminish with increasing $\omega$ or $\gamma$. These findings are consistent with those reported in Ref. \cite{Ladino:2022aja}. In fact, our analysis of the constraint relations among $\omega,\,\gamma,$ and $a$ already revealed that the effect introduced by $G(r)$ can be interpreted in terms of an effective mass—specifically, the effective mass decreases as $\omega$ and $\gamma$ increase. As a result, not only do the radii of the MBO and ISCO shrink, but the radius of the event horizon also decreases accordingly as shown in Fig. \ref{fig:r0vsgamma}.

Based on the above results, we further investigated how the gravitational waveforms emitted from periodic orbits are influenced by $\omega$ and $\gamma$. The results indicate that in the case of prograde orbits, the maximum orbital turning point increases with $\omega$ and $\gamma$, and the deviation of the waveforms from that in the classical Kerr background also becomes more pronounced as $\omega$ and $\gamma$ increase. These findings are consistent with the results presented in Sec. \ref{sec:timelike}. The reduction in effective mass inevitably weakens the gravitational binding of the RGI-Kerr black hole on the secondary objects. In contrast, for retrograde orbits, under a fixed angular momentum, the variation in orbital energy induced by different values of $\omega$ and $\gamma$ is much smaller compared to that in prograde cases. As a result, the gravitational waveform is less affected by changes in $\omega$ and $\gamma$. This can be attributed to the fact that retrograde periodic orbits require higher energy and angular momentum for timelike particles, making them less sensitive to parameter variations.

These results illustrate the dominant influence of renormalization group corrections—specifically the cutoff parameters $\omega$ and $\gamma$—on both the timelike particle geodesics and the gravitational waves emitted by periodic orbits under the geodesic approximation in the RGI-Kerr spacetime. They provide valuable insights for further understanding the effectiveness of the asymptotic safety approach to quantum gravity, and may also serve as a reference for studying black hole backgrounds derived through alternative methods within other asymptotic safe gravity frameworks. Furthermore, our analysis suggests that the characteristic frequencies of gravitational radiation from EMRIs in periodic orbits typically lie within the frequency bands $10^{-3}$ Hz $\sim$ $0.1$ Hz, which is sensitive to LISA, eLISA, Taiji, TianQin and DECIGO. Especially, the corresponding characteristic strains exceed the sensitivity curve of DECIGO, but it is anticipated that more gravitational wave detectors in the future may be able to observe the gravitational wave signals emitted by EMRIs in the RGI-Kerr black hole background. 

Our current investigation is limited into the adiabatic approximation in which the gravitational wave’s backreaction to the particle’s motion is ignored for easing our analysis. This consideration is sufficient since we only consider one complete period of the orbital motion. {Noted that the quantum parameters in the current model modify the EMRI dynamic at small correction, so their imprint on GW may challenge to be detected. Besides, there could be potential degeneracies associated with modified gravity effects or astrophysical environments such as accretion disks. However, considering that the parameter degeneracy is a common and significant challenge in observational test, the multi-messenger or multi-band observations and more comprehensive waveform families to testify the degeneracies deserve further studies. Therefore, one natural direction is to explore the possible effect of gravitational radiation on the evolution of periodic orbits, and also generate more accurate waveform to detect the renormalization group corrections by the observation of an EMRI by future space-based gravitational wave detector \cite{Maselli:2021men}. Moreover, connecting the observations related with the current EMRI Gw,  black hole imagings \cite{EventHorizonTelescope:2019dse,EventHorizonTelescope:2022wkp} and the dynamic of S2 orbit in supermassive black hole \cite{Do:2019txf} to probe the quantum parameters could be beneficial for testifying their potential effect and degeneracies.} We hope to address these issues in the future.

\begin{acknowledgments}
This work is partly supported by Natural Science Foundation of China under Grants No.12375054. YZ is
also supported by the research funds No. 2055072312.
\end{acknowledgments}

\nocite{*}
\bibliography{ref}
\bibliographystyle{utphys}

\end{document}